\documentstyle[12pt]{article}
\parskip=\medskipamount

\title{Classical description of spinning degrees of freedom of 
relativistic particles  by means of commuting spinors.}
\author{A.A. Deriglazov\thanks{deriglaz@fma.if.usp.br ~ On leave of 
absence from Dept. Math. Phys., Tomsk Polytechnical University, Tomsk, 
Russia}
~ and ~ D.M. Gitman\thanks{gitman@fma.if.usp.br}}
\date{Instituto de F\'\i sica, Universidade de S\~ao Paulo,\\
P.O. Box 66318, 05315-970, S\~ao Paulo, SP, Brasil.}

\begin{document}
\maketitle
\large
\begin{abstract}
We consider a possibility  to describe spin one-half and higher 
spins of  massive relativistic particles by means of commuting spinors. 
We present two classical  gauge  models with the variables  
$x^\mu,\xi_\alpha,\chi_\alpha$, where $\xi,\chi$
are commuting Majorana spinors. In course of quantization both 
models reproduce Dirac equation. We analyze the possibility to introduce an 
interaction with an external electromagnetic background into the models 
and to generalize them to higher spin description. The first model  
admits a minimal 
interaction with the external electromagnetic field, but leads 
to reducible representations of the Poincare group being generalized for 
higher spins. The second model turns out to be appropriate for description 
of the massive higher spins. However, it seams to be difficult to  introduce a 
minimal interaction with an external electromagnetic field into this model. 
We  compare our approach with one, which uses Grassman variables,  and 
establish a  relation between them.

\end{abstract}
\noindent
{\bf PAC codes:} 0370\\
{\bf Keywords:} Quantization, higher spins, classical models.

\section{Introduction.}

Classical relativistic spinning particle models  and their quantization have
been discussed intensively for a long time [1-13]. Such   models have to
 reproduce one particle sector of the corresponding quantum field theory
 in course of quantization. There are two different approaches for 
description of the spinning degrees of freedom in such models. In the 
first one the anticommuting (Grassman)
variables are used [2-13] following to the pioneer works [2-6]. 
In the second approach one uses commuting variables, which
parametrize some compact manifolds \cite{1,14,15}. Both approaches have some
advantages and problems. In particular, it turns out to be problematic
to generalize the first  approach in $D=3+1$ dimensions 
to massive higher spins, and to introduce 
an  interaction with external backgrounds  for the case
of higher spins. The last problem appears also in the second
approach. 

In the present work we consider a possibilitiey  to describe spin 
one-half and higher spins of  massive relativistic particles by means of 
commuting spinors. We present two classical  gauge  models in $D=3+1$ 
dimensions
 with the variables  $x^\mu,\xi_\alpha,\chi_\alpha$, where $\xi,\chi$
are commuting Majorana spinors. Both  models are obtained by means of a 
localization of some global symmetries, which are characteristic for
 a simple 
action  containing only  kinetic terms for the above variables. In course 
of quantization both models reproduce Dirac equation.
We analyze a 
possibility to introduce an 
interaction with an external background into the models and to generalize them
to higher spin description. The first model  admits a minimal 
interaction with the external electromagnetic field, but leads 
to reducible representations of the Poincare group being generalized for 
higher spins. The second model turns out to be appropriate for the 
description of the massive higher spins,it leads to  Bargmann-Wigner wave
equations \cite{16} in  course of quantization. However, it
seams to be difficult to  
introduce a minimal interaction with an external electromagnetic field into 
this model. We  compare our approach with one, which uses 
Grassmann 
variables,  and establish a  relation between them. In particular, 
to this end we discuss  a possibility to generalize basic models with
Grassman variables to the case of massive higher spins

\section{Description of the spinning degrees of freedom in terms of
  Grassman variables. Problem with higher spins.}

In the most symmetric form, the action for spin
one-half particle in $D=3+1$ can be written as  [2-6]
\begin{equation}\label{1}
S =\int_0^1 \left[
  \frac 1{2e}(\dot x^\mu-i\chi\psi^\mu)^2  - e\frac{m^2}{2}  
+i m\psi^5 \chi +i\psi_n\dot{\psi}^n   \right]d\,\tau\, ,
\end{equation}
where $x^\mu$, $e$ are ordinary (bosonic or even) variables and 
$\psi^n$, $\chi$ are Grassman (fermionic or odd) variables which 
describe spinning degrees of freedom. Greek indices
run over $\overline{0,3}$ and Latin ones $n,m$ run over $\overline{0,3},5$.
The metric tensors:
$\eta_{\mu\nu}= {\rm diag}(-1+1+1+1)$ and 
$\eta_{mn}= {\rm diag}(-1+1+1+1+1)$. 
In addition to the reparametrizations, the action is invariant under 
local $N=1$ (world-line) supersymmetry transformations.
 
In the Hamiltonian formulation there are the following constraints 
\begin{eqnarray}\label{2}
&&P_e = \frac{\partial L}{\partial\dot{e}} = 0, ~ 
P_\chi = \frac{\partial_r L}{\partial\dot{\chi}} = 0, ~ 
P_n -i\psi_n=0, ~ 
\left(P_n =  \frac{\partial_r L}{\partial\dot{\psi}^n}\right), \nonumber \\
&&p_\mu\psi^\mu +m\psi^5 =0, ~ 
p^2 + m^2=0, ~ 
\left(p_\mu = \frac{\partial L}{\partial\dot{x}^\mu}\right).
\end{eqnarray}
The Dirac brackets are defined by means of the second-class
constraints $P_n -i\psi_n=0$.
 For the variables $\psi^n$ these brackets are
\begin{equation}\label{3}
\{\psi^n, \psi^m\}_D=-i\eta^{nm},
\end{equation}
and define commutation relations for the corresponding 
operators $\hat\psi^\mu$,
\begin{equation}\label{4}
\left[\hat\psi^n, \hat\psi^m\right]_+=\hbar\eta^{nm},
\end{equation}
The Dirac brackets for the remaining variables coincide with the
Poisson ones. The ommutation relations for operators $\hat\psi$ can
be realized in a space of four-dimensional columns $\Psi_\alpha$ as follows:
\begin{eqnarray}\label{5}
\hat\psi^\mu=\left(\frac {\hbar}2\right)^{\frac 12}
\Gamma^5\Gamma^\mu, \quad
\hat{\psi}^5=\left(\frac {\hbar}2\right)^{\frac 12}\Gamma^5,
\end{eqnarray}
where   $\Gamma^\mu$ are $\gamma$-matrices in $D=3+1$ dimensions and 
$\Gamma^5=i\Gamma^0\Gamma^1\Gamma^2\Gamma^3$. Applying operators of 
 the first-class constraints to state vectors, we  specify according to 
Dirac \cite{28} the physical sector. It follows from the structure of 
the first-class constraints that the physical sector 
contains only 
vectors of the form $\Psi_\alpha (x^\mu)$ subjected to the Dirac equation 
\begin{eqnarray}\label{6}
(\hat p_\mu\Gamma^\mu+m)\Psi=0.
\end{eqnarray}
Due to the fact that the Hamiltonian is zero in the case under consideration, 
no more equations on state vectors appear.
Thus, the model (\ref{1})reproduces the Dirac equation in course of such 
simplified quantization. One can show that the consistent 
canonical quantization leads to the same result \cite{7}.

The basic model (\ref{1}) admits a natural introduction of an interaction
with electromagnetic and gravitational backgrounds [2-6]. The
limit $m \to 0$ was studied in \cite{17,18} (actions, which may 
describe  Weyl particles, were considered in \cite{9,19,20,32}).
A generalization of the action (\ref{1}) to arbitrary even $D=2n$ dimensions
turned out to be trivial \cite{17}, whereas the generalization to an
odd dimensions $D=2n+1$ met complication due to the absence of
$\Gamma^5$ in such a case. Different ways of the problem solution were
proposed in \cite{21,22,23} for the case  $D=2+1$. One of the corresponding 
actions  was generalized then  \cite{24} to
arbitrary $D=2n+1$ dimensions. 

Construction of models for  relativistic particles with   higher spin 
turn out to be nontrivial and seams to be in general case an open
problem until now. Using the basic action
(1) one can try to construct a  model in $D=3+1$ which
describes massive spin $s=N/2$
by means of an extension of the odd sector as
$\psi^n_a, \chi_a, ~ a=1, \cdots ,N$ \cite{11,12,8} 
\begin{eqnarray}\label{7}
&&S=\int d\tau\left\{\frac 1{2e}(\dot x^\mu-i\chi_a\psi^\mu_a)^2
    -e\frac{m^2}{2}    \right. \nonumber \\
&&\left.+im\psi^5_a\chi_a  +  i\psi_{an}\dot\psi^n_a      +
\frac{i}2f_{ab}(\psi_{an}\psi^n_b+k_{ab})\right\}.
\end{eqnarray}
The action (\ref{7}) is invariant under reparametrizations, 
local $N$-extended
supersymmetry transormations and local $O(N)$-transformations.
The Chern-Simons term $k_{ab}f_{ab}$ can only  be added
in the case $N=2$ without breaking of $O(N)$ symmetry.
Hamiltonian analysis for the action (\ref{7}) leads to the 
following essential first-class constraints 
\begin{eqnarray}
&&\psi^\mu_{[a}\psi^\mu_{b]}+\psi^5_{[a}\psi^5_{b]}+k_{ab}=0, \label{8}\\
&&p_\mu\psi^\mu_a+m\psi^5_a=0, \label{9} \\
&&p^2+m^2=0. \label{10}
\end{eqnarray}
and to  Dirac brackets for the spinning degrees of freedom
\begin{eqnarray}\label{11}
\{\psi^n_a, \psi^m_b\}_D=-i\eta^{nm}\delta_{ab}\;.
\end{eqnarray}

Let us consider first the case $k_{ab}=0$ in the action. 
To analyze this case it is convenient to impose 
the following gauge conditions
\begin{eqnarray}\label{12}
\psi^0_a=0,
\end{eqnarray}
for the first-class constraints (\ref{9}). Then due to the constraint
(9) and to the gauge condition (12), one stays with a set of three
independent Grassman variables, which is convenient to select as
\cite{18},  
\begin{eqnarray}\label{13}
\theta^i_a=\psi^i_a-\frac{p^i}{ {\bf p}^2}p_i\psi^i_a +
\frac{ p^i}{ {\bf p}^2}({\bf p}^2+m^2)^{\frac 12}\psi^5_a, 
\end{eqnarray}
The Dirac brackets for the variables $\theta^î_a$ are
\begin{eqnarray}\label{14}
\{\theta^i_a , ~ \theta^j_b\}_D=-i\eta^{ij}\delta_{ab},
\end{eqnarray}
while the first-class constraints (\ref{8}) take a form 
\begin{eqnarray}\label{15}
\theta^i_{[a}\theta^i_{b]}=0.
\end{eqnarray}
Operators, which correspond to the variables 
$\theta^i_a$, can be realized as follows
\begin{eqnarray}\label{16}
\hat\theta^i_a=\left(\frac{\hbar}2\right)^{\frac 12}\Gamma^5\otimes
\cdots \otimes\Gamma^5\otimes
\Gamma^5\Gamma^i_{(a)}\otimes 1\otimes\cdots\otimes 1,
\end{eqnarray}
in a space of spin-tensor functions $\Psi_{\alpha_1 {\dots}
  \alpha_N} (x^\mu)$.
Applying the operators of the first-class
  constraints (\ref{15}) to state vectors and using  Fierz identities, 
we get  the conditions 
\begin{eqnarray}\label{17}
(\Gamma_i\Gamma^{(k)}\Gamma^i)_{\gamma\beta}
\Psi_{\alpha_1\cdots\beta\cdots\gamma\cdots\alpha_N}=0,
\end{eqnarray}
where $\Gamma^{(k)}\equiv\{1, \Gamma^\mu, \Gamma^{\mu\nu},
\Gamma^\mu\Gamma^5, \Gamma^5\}$ is a basis in the 
space of $4\times 4$ matrices.
The equations (\ref{17}) are
nontrivial for any $k=0,1,2,3,4$. One can find by straightforward 
calculations that $\Psi_{\alpha_1 {\dots} \alpha_N}=0$ as a consequence 
of (\ref{17}), i.e. the physical subspace is empty. Thus,
the direct generalization of the basic action (\ref{1}) to higher spin
massive case turns out to be problematic.

The modified action (\ref{7}), which contains the Chern-Simons term 
$(k_{ab}\ne 0)$, leads to the constraints
\begin{eqnarray}\label{18}
\theta^i_{[a}\theta^i_{b]}+k_{ab}=0,
\end{eqnarray}
instead of (\ref{15}). For $N>2$ they are  mixture of first and second-class
constraints. The first-class constraints, being separated from the
second-class ones, have the form (\ref{15}), that immediately leads 
to the empty physical subspace. 
The case $N=2$ is an exceptional since the rotational
symmetry is not broken and the constraint (\ref{18}) turns out to be 
first-class. It was shown in \cite{18} that the canonical quantization of 
such a model
reproduces adequate quantum  description of spin one massive particle.

The massless case can be obtained from (\ref{7}-\ref{11}) by the
 substitution $m=0, \psi^5_a=0, k_{ab}=0$. The model leads to 
 Bargmann-Wigner equations for massless spin $s=N/2$ in course of the 
quantization,  i.e. to  the irreducible 
representation of the complete Poincare group.  
All the pseudoclassical constructions in $D=3+1$
case may be extended to any even dimensions $D=2n$. 
To construct higher spin models in $D=2n+1$ one may start with a model
for $s=1/2$ in the same dimension \cite{25,26}. For $D>2+1$ a detailed
elaboration of such  a way seams to be still an open
problem. Introduction of the interaction with external backgrounds
remains also unsolved problem for higher spin pseudoclassical models. 

Existence of the above mentioned problems in pseudoclassical approach
motivates a development of alternative descriptions of spinning degrees of
freedom in terms of the bosonic variables. Below we present two new
models of such a kind.

\section{First model. Spin one-half from the commuting spinors.}

We start from a reparametrization invariant action of spinless 
relativistic particle  in $D=3+1$ dimensions  and  add to it a simplest 
reparametrization invariant term, which may be constructed from two
additional (to $x^\mu$ and $e$) variables 
$\chi_\alpha$ and $\xi_\beta,\;\alpha,\beta=1,2,3,4,$, the latter are  
commuting Majorana spinors of SO(1,3) group,
\begin{eqnarray}\label{19}
S=\int d\tau\left\{\frac 1{2e}\dot x^2-e\frac{m^2}{2}+
i\bar\chi\dot\xi\right\}\;.
\end{eqnarray}
Besides of the manifest Poincare invariance, the action (\ref{19}) is also
invariant under global symmetry transformations with 
parameters $\beta, \gamma, \epsilon$, where $\beta,\gamma$ are scalars,
while $\epsilon$ is a commuting spinor,
\begin{eqnarray}
&&\delta\xi=-\beta\xi, \qquad
\delta\bar\chi=\beta\bar\chi, \label{20} \\
&&\delta\xi=-\frac{\gamma}{e}\dot x_\mu\Gamma^\mu\xi, \qquad
\delta\bar\chi=\frac{\gamma}{e}\dot x_\mu\bar\chi\Gamma^\mu, \qquad
\delta x^\mu=i\gamma(\bar\chi\Gamma^\mu\xi), \label{21} \\
&&\delta\bar\xi=\frac 1{e}\dot x_\mu\bar\epsilon\Gamma^\mu, \qquad
\delta x^\mu=-i\bar\epsilon\Gamma^\mu\chi. \label{22}
\end{eqnarray}
Let us consider some local versions of the theory (\ref{19}) which can
be obtained by means of gauging the 
symmetries (\ref{20}), (\ref{21}), (\ref{22}). First, we consider a
model which arises after localization of the transformations (\ref{20}),
(\ref{21}). Following the usual manner, one has to consider the
parameters $\beta, \gamma$ as  arbitrary functions of the
evolution parameter $\tau$ and to introduce a ``covariant'' derivative,
\begin{eqnarray}\label{23}
D\xi\equiv\dot\xi+\phi\xi, \qquad
\delta\phi=-\dot\beta,
\end{eqnarray}
for the symmetry (\ref{20}), and another one
\begin{eqnarray}\label{24}
Dx^\mu\equiv\dot x^\mu-i\omega(\bar\xi\Gamma^\mu\chi),\qquad
\delta\omega=\dot\gamma,
\end{eqnarray}
 for the symmetry (\ref{21}). The new variables $\phi, \omega$ play a role
of ``gauge fields'' for the corresponding symmetries, as it can be seen
from their transformation low. Besides, one can add a terms $i k \phi$,
$i k_1 m \omega$ (where $k$ and $k_1$ are some constants to be
specified below), which does not break both the reparametrization symmetry and 
 $\beta(\tau)$ , $\gamma(\tau)$ symmetries. Thus, a local
version for the model (\ref{19})) can be written as follows
\begin{eqnarray}
&&S=\int d\tau\left\{\frac 1{2e}Dx^\mu Dx_\mu+i\bar\chi D\xi-
e\frac{m^2}{2}+ik_1m\omega+ik\phi\right\}, \label{25} \\
&&\delta\xi=-\beta\xi, \qquad
\delta\bar\chi=\beta\bar\chi, \qquad
\delta\phi=-\dot\beta \label{26} \\
&&\delta\xi=-\frac{\gamma}{e}\dot x_\mu\Gamma^\mu\xi, \;\;
\delta\bar\chi=\frac{\gamma}{e}\dot x_\mu\bar\chi\Gamma^\mu,\;\;
\delta x^\mu=i\gamma(\bar\chi\Gamma^\mu\xi), \,\;
\delta\omega=\dot\gamma. \label{27} 
\end{eqnarray}
The action can be written also in the first order form
\begin{eqnarray}\label{28}
&&S=\int d\tau\left\{\pi_\mu\dot x^\mu-
\frac12 e(\pi^2+m^2)
+i\bar\chi\dot\xi
-i\omega(\pi_\mu\bar\xi\Gamma^\mu\chi-k_{1}m)-\right.\nonumber \\
&&\left.i\phi(\bar\xi\chi-k)\right\}.
\end{eqnarray}
It shows explicitly the structure of secondary first-class constraints, which
correspond to the symmetries (\ref{26}), (\ref{27}), with 
$\omega, \phi$ being considered as the Lagrange  multipliers.

In the Hamiltonian formulation \cite{28,29}, a complete set of 
constraints for the  
action (\ref{28}) under consideration has the form
\begin{eqnarray}
&&p_e=p_\omega=p_\phi=0, \label{29} \\
&&p^\mu-\pi^\mu=0,\;\;
p^\mu_\pi=0,\;\;
\bar p_\xi-i\bar\chi=0,\;\; 
\bar p_\chi=0, \label{30} \\
&&p_\mu\bar\xi\Gamma^\mu\chi-k_1m=0,\;\; 
p^2+m^2=0,\;\;
\bar\xi\chi-k=0, \label{31}
\end{eqnarray}
where $p,p_\pi,\bar p_\xi,\cdots$ are canonical momenta for the
variables $x,\pi,\xi,\cdots$, and the equations (\ref{31}) represent
secondary constraints. There are no more constraints in the problem.  

Imposing the gauge conditions
\begin{eqnarray}\label{32}
e=1, \qquad \phi=\omega=0,
\end{eqnarray}
we define the Dirac bracket associated with the second-class set
(\ref{29}), (\ref{30}), (\ref{32}). For the independent
variables $x^\mu, p_\nu ,\bar\xi, \chi$ they are
\begin{eqnarray}\label{33}
\{x^\mu, p_\nu\}_D=\delta^\mu{}_\nu, \qquad
\{\bar\xi^\alpha, \chi_\beta\}_D=-i\delta^\alpha{}_\beta.
\end{eqnarray}
One can see that in the gauge chosen, the variables obey the free
equations of motion: 
$\dot x^\mu=p^\mu, \dot p^\mu=0, \dot{\bar\xi}=\dot\chi=0$, 
and the first-class constraints (\ref{31}).

The Dirac brackets (\ref{33}) define the commutation relations for the
corresponding operators
\begin{eqnarray}\label{34}
[\hat x^\mu,\hat p_\nu]_{-}=i\hbar\delta^\mu{}_\nu, \qquad
[\hat{\bar\xi^\alpha},\hat\chi_\beta]_{-}=\hbar\delta^\alpha{}_\beta.
\end{eqnarray}
The algebra (\ref{34}) can be realized on the space of $\bar\xi$-regular
functions
\begin{eqnarray}\label{35}
\Psi(x,\bar\xi)=\sum_{N=0}^\infty\Psi^{(N)}\equiv\sum_{N=0}^\infty
\bar\xi^{\alpha_1}\cdots\bar\xi^{\alpha_N}
\Psi_{\alpha_1\cdots\alpha_N}(x^\mu),
\end{eqnarray}
where $\Psi_{\alpha_1\cdots\alpha_N}$ is spin-tensor of N-order. By
the construction, it is symmetric in all its indexes: 
$\Psi_{\alpha_1\cdots\alpha_N}\equiv\Psi_{(\alpha_1\cdots\alpha_N)}$.
We select the standard coordinate realization
\begin{eqnarray}\label{36}
\hat x^\mu=x^\mu, \quad
\hat p_\mu=-i\hbar\partial_\mu, \quad 
\hat{\bar\xi^\alpha}=\bar\xi^\alpha, \quad
\hat\chi_\alpha=-\hbar\frac{\partial}{\partial\bar\xi^\alpha}\,,
\end{eqnarray}
and specify the physical sector applying operators of  first-class
constraints (\ref{31}) to state vectors. Let us choose $k_1=k=\hbar$ 
in the initial action (\ref{25}),
then the last constraint from (\ref{31}) leads to the conditions 
\begin{eqnarray}\label{37}
(N-1)\Psi^{(N)}=0, \qquad
N=0,1,\cdots,
\end{eqnarray}
which means that only the subspace
$\Psi^{(1)}=\bar\xi^\alpha\Psi_\alpha(x)$ from $\Psi$-space
(\ref{35}) contains physical vectors. Then
the remaining constraints from (\ref{31}) demand that  
vectors $\Psi_\alpha(x)$ must obey
the Dirac equation
\begin{eqnarray}\label{38}
(\hat p_\mu\Gamma^\mu+m)\Psi=0\;.
\end{eqnarray}
Thus, under an appropriate choice of the parameters, the action (\ref{28})
describes spin one-half free particle. 

An interaction with an external electromagnetic field $A^\mu$, 
may be introduced by means of a minimal coupling,
\begin{eqnarray}\label{39}
S_{int}=\int d\tau\left\{qA_\mu\dot x^\mu-
\frac 12eqF_{\mu\nu}(\bar\xi\Gamma^{\mu\nu}\chi)\right\},
\end{eqnarray}
where $q$ is an electric charge and the coefficient 
$-\frac 12$ in the second term was fixed from the requirement of 
invariance of (\ref{39}) under the transformations
(\ref{27}). Repeating the above Hamiltonian analysis, one gets the 
equations of motion 
\begin{eqnarray}\label{40}
&&\dot x^\mu=\tilde p^\mu, \qquad
\dot p^\mu=-\frac 12q\partial^\mu
F_{\nu\rho}(\bar\xi\Gamma^{\nu\rho}\chi), \nonumber \\
&&\dot{\bar\xi}=0, \quad
\dot\chi=-\frac{i}2qF_{\mu\nu}\Gamma^{\mu\nu}\chi\;,
\end{eqnarray}
and the constraints 
\begin{eqnarray}\label{41}
\tilde p^2+m^2+qF_{\mu\nu}(\bar\xi\Gamma^{\mu\nu}\chi)=0, \quad
\tilde p_\mu(\bar\xi\Gamma^\mu\chi)-\hbar m=0, \quad
\bar\xi\chi-\hbar=0,
\end{eqnarray}
where $\tilde p^\mu=p^\mu-qA^\mu$. Being considered as
constraints on the state vectors (\ref{35}), the equations (\ref{41}) 
reproduce formally a description of spin one-half particle on an external
electromagnetic field. 

It follows from (\ref{37}) that in the realization under consideration
the last constraint from (\ref{31}) plays a role of the second Casimir 
operator for the Poincare group. Thus, it is interesting to consider 
the case when
$k_1=k=\hbar N$ in the initial action. Then the equation (\ref{37}) 
specifies the subspace $\Psi^{(N)}=\bar\xi^{\alpha_1}\cdots\bar\xi^{\alpha_N}
\Psi_{\alpha_1\cdots\alpha_N}(x)$ from $\Psi$-space (\ref{35}), 
while the first two constraints from (\ref{31}) lead to the conditions
\begin{eqnarray}\label{42}
&&(p_\mu\Gamma^\mu_{\alpha_1\beta}+m\delta_{\alpha_1\beta})
\Psi_{\beta\alpha_2\cdots\alpha_N}+
(p_\mu\Gamma^\mu_{\alpha_2\beta}+m\delta_{\alpha_2\beta})
\Psi_{\alpha_1\beta\alpha_3\cdots\alpha_N}+\cdots=0\;, \nonumber \\
&&(p^2+m^2)\Psi_{\alpha_1\cdots\alpha_N}=0\;.
\end{eqnarray}
The subspace which is defined by (\ref{42}) contains, in particular, 
an irreducible representation of spin $s=N/2$,
\begin{eqnarray}\label{43}
(p_\mu\Gamma^\mu_{\alpha_k\beta}+m\delta_{\alpha_k\beta})
\Psi_{\alpha_1\cdots\beta\cdots\alpha_N}=0, \quad
k=1,2,\cdots,N.
\end{eqnarray}
It would be interesting to study in more detail the spin content of the
space (\ref{42}). 

Thus, the action (\ref{28}) with commuting spinor variables
reproduces after a quantization an adequate quantum theory of  spin 
one-half particle. However, it seems to be not appropriate  for minimal 
description of higher spins. In the next section
we are going to consider another action, which solves the latter 
problem. This action  corresponds to
localization of symmetries (\ref{20}), (\ref{22}) instead 
of (\ref{20}), (\ref{21}). 

\section{Second model. Higher spins from the commuting spinors.}

Action functional to be examined is
\begin{eqnarray}\label{44}
S=\int d\tau\left\{\pi_\mu\dot x^\mu-\frac
  12e(\pi^2+m^2)+i\bar\chi\dot\xi- \right.\cr
\left.
  i\bar\lambda(\pi_\mu\Gamma^\mu\chi-k_1m\chi)-i\phi(W^2-k)\right\}\;,
\end{eqnarray}
where $\bar\lambda^\alpha$ is an additional commuting Majorana spinor, 
$W_\mu\equiv\frac 12\epsilon_{\mu\nu\rho\delta}S^{\nu\rho}\pi^\delta$
is the Pauli-Lubanski vector, and 
$S^{\nu\rho}=-\frac 12\bar\xi\Gamma^{\nu\rho}\chi$.
The latter quantity is a generator of Lorentz transformations, which
are induced by the Dirac brackets (\ref{33}):
$\delta_\omega\bar\xi=\omega_{\mu\nu}\{S^{\mu\nu},\bar\xi\}_D=
-\frac{i}2\omega_{\mu\nu}\bar\xi\Gamma^{\mu\nu}$.

Let us demonstrate that the action (\ref{44}) can be considered as a local
version of the model (\ref{19}). For this aim we rewrite (\ref{44}) 
in the second
order form by means of an integration over the variable $\pi^\mu$. First,
by using of $\gamma$-matrix identities as well as the identity
\begin{eqnarray}\label{45}
\epsilon^{dabc}\epsilon_{d\mu\nu\rho}=-\delta^a_\mu
(\delta^b_\nu\delta^c_\rho-\delta^c_\nu\delta^b_\rho)+ \cr
\delta^b_\mu(\delta^a_\nu\delta^c_\rho-\delta^c_\nu\delta^a_\rho)-
\delta^c_\mu(\delta^a_\nu\delta^b_\rho-\delta^b_\nu\delta^a_\rho),
\end{eqnarray}
the second Casimir operator $W^2$ can be  rewritten as
\begin{eqnarray}\label{46}
W^2=-\frac 18\pi^2(\bar\chi\Gamma^{\mu\nu}\xi)(\bar\chi\Gamma_{\mu\nu}\xi)+
\frac
14(\bar\chi\Gamma^{\mu\rho}\xi)(\bar\chi\Gamma_{\mu\nu}\xi)p_\rho p^\nu \cr
=-\frac 1{16}\pi^2(\bar\xi\chi)^2+
\frac 14\pi_\mu(\bar\xi\bar\sigma^\mu\chi) 
\xi(\pi_\nu\sigma^\nu\bar\chi-m\chi)+ \cr
\frac{m}4(\bar\xi\chi)\bar\xi(\pi_\mu\bar\sigma^\mu\chi-m\chi)+
\frac 14(\bar\xi\bar\chi)(\xi\chi)(\pi^2+m^2),
\end{eqnarray}
where the last three terms are written in two
dimensional spinor notations, and
$\sigma^\mu{}_{b\dot a},\bar\sigma^{\mu\dot a b}$
are $D=4$ matrices Pauli \cite{27}. After substitution of (\ref{46}) 
into (\ref{44}), the
last three terms can be included into redefinition of the variables 
$e,\lambda^a,\bar\lambda_{\dot a}$. As a result, one obtains
the action in the form 
\begin{eqnarray}\label{47}
S=\int d\tau\left\{\pi_\mu\dot x^\mu+i\bar\chi\dot\xi-
\frac 12e(\pi^2+m^2)+ \right. \cr
\left. \frac {i}{16}\phi\pi^2(\bar\xi\chi)^2-i\bar\lambda
(\pi_\mu\Gamma^\mu\chi-k_1m\chi)+ik\phi\right\}.
\end{eqnarray} 
Equations of motion ${\delta S}/{\delta\pi_\mu}=0$ can be solved as
\begin{eqnarray}\label{48}
\pi^\mu=\frac{\dot x^\mu-i\bar\lambda\Gamma^\mu\chi}
{e-\frac{i}8\phi(\bar\xi\chi)^2},n
\end{eqnarray}
and substituted into (\ref{47}). After a additional redefinition 
of the variable $e: e\to e+\frac{i}8\phi(\bar\xi\chi)^2$,
the action takes the final form
\begin{eqnarray}\label{49}
S=\int d\tau\left\{\frac 1{2e}Dx^\mu Dx_\mu+i\bar\chi D\xi
-\frac 12 em^2+ \right. \cr
\left.ik_1m\bar\lambda\chi+ik\phi\right\}\;,
\end{eqnarray}
where 
$Dx^\mu\equiv\dot x^\mu-i\bar\lambda\Gamma^\mu\chi$,
$D\xi\equiv\dot\xi-\frac 1{16}\phi m^2(\bar\chi\xi)\xi$. Local
symmetries of the action are both reparametrizations and the 
transformations 
\begin{eqnarray}
&&\delta\xi=-\beta\xi,\; 
\delta\bar\chi=\beta\bar\chi,\; 
\delta\phi=-\frac{16}{m^2(\bar\chi\xi)}\dot\beta\;, \label{50} \\
&&\delta\bar\xi=\bar\epsilon(\frac 1{e}Dx_\mu\Gamma^\mu-m),\;\; 
\delta x^\mu=-i\bar\epsilon\Gamma^\mu\chi, \;\;
\delta\bar\lambda=-\dot{\bar\epsilon}+
\frac{m^2}8(\bar\chi\xi)\bar\epsilon\;, \label{51}
\end{eqnarray}
with the parameters $\beta(\tau)$, $\epsilon_\alpha(\tau)$. Taking
into account that
the combination $(\bar\chi\xi)$ is invariant under the transformations
(\ref{50}), and comparing (\ref{50}), (\ref{51}) with 
the equations (\ref{20}), (\ref{22}), one may see that the action 
(\ref{49}) can be obtained from (\ref{19}) by
localization of the symmetries (\ref{20}), (\ref{22}). The variables
$\phi,\lambda_\alpha$ play a role of the corresponding gauge
fields. It is interesting to remark that if one starts from the action
(\ref{19}) with the anticommuting spinor variables $\xi,\chi$, the 
analogous procedure leads to the model of Siegel 
superparticle \cite{30,31}.

Hamiltonian analysis for the action (\ref{44})  leads to the following 
 first-class constraints 
\begin{eqnarray}
&&\frac 1{16}(\epsilon_{\mu\nu\rho\delta}(\bar\xi\Gamma^{\nu\rho}\chi)
p^\delta)^2-k=0, \label{52} \\
&&p^2+m^2=0,\;\; 
(p_\mu\Gamma^\mu-k_1m)\chi=0, \label{53}
\end{eqnarray}
while the remaining variables $x^\mu,p_\nu,\bar\xi,\chi$ obey the
Dirac brackets (\ref{33}). Commutation relations for the corresponding
operators (\ref{34}) can be realized similar to  
(\ref{35}), (\ref{36}). After tedious calculations, one may see that 
the first-class constraint (\ref{52}) leads to the following condition 
on the physical states
\begin{eqnarray}\label{54}
\left[\frac 1{16}\hbar^2m^2N(N+2)-k\right]\Psi^{(N)}=0, \quad
N=0,1,2,\cdots.
\end{eqnarray}
Let us take $k=\frac 1{16}\hbar^2m^2N(N+2), ~ k_1=-1$ in the 
action (\ref{49}). Then the condition (\ref{54}) selects the 
subspace $\Psi^{(N)}$ 
from (\ref{35}) as physical one, while the constraints (\ref{53}) leads to
 the Bargmann-Wigner equations (\ref{43}) for the functions 
 $\Psi_{(\alpha_1\cdots\alpha_N)}(x)$, which describe  spin $s=N/2$ 
relativistic particles. 

Thus, the action (\ref{44}) provides classical description of massive 
higher spin relativistic 
particles and leads to an adequate minimal quantum theory of these 
particles.  However, the problem 
how  to introduce an interaction with an external electromagnetic 
background is still open for the 
 case under consideration since the minimal interaction does not
retain the first-class character of the constraints (\ref{53}).

\section{Discussion.}

Thus, it was  demonstrated that two simple classical models (\ref{25}),
(\ref{44}), in which  commuting spinors are used to describe spinning 
degrees of freedom, lead to adequate 
quantum description of spinning particles, in particular,   irreducible 
representations of the complete Poincare group in the Bargmann-Wigner 
realization (\ref{43}) are selected in course of the quantization.

One may mention some advantages of using the commuting spinors.
In course of the quantization we get directly a realization in which 
 state vectors are symmetric in all their spinor indexes by construction. 
As it was demonstrated above, in  the models, based on  odd variables,  an 
additional O(N) local symmetry has to be introduced   to provide such a 
realization. Beside the mass-shell condition, the second Casimir 
operator (\ref{46}) can be naturally incorporated into the action \cite{15}, 
which allows one to
define a subspace of a given spin. For models with  odd
variables the classical analog of this operator is identically zero
for $D=4$. 

However, the approach proposed has not be treated as an alternative to the 
pseudoclassical description of spinning particles. It has to be 
rather considered as a combination of classical description of space motion 
with semi-classical description of spin. Namely, the models proposed may be 
understood as those which where obtained from some pseudoclassical models 
by a partial quantization of Grassman variables. Indeed, let us take the 
action (\ref{1}) and quantize only the odd variables $\psi$, 
considering $x^\mu$ and $e$ as external given fields. Then we arrive to the 
$\gamma$-matrix realization for the operators $\hat{\psi}$. In this case, 
the first-class constraint being applied to state vectors coincides with 
the classical equation of motion (\ref{53}) 
of the theory (\ref{44}). That 
confirms the above interpretation of the status of the models proposed.

\section*{Acknowledgments.}

D.M.G. thanks Brazilian foundation CNPq for permanent support. The
work of A.A.D. has been supported by the Joint DFG-RFBR project No
96-02-00180G, by Project INTAS-96-0308, and by FAPESP.

\end{document}